\documentclass[11pt]{article}

\newcommand{\up}[1]{^{\mathrm{#1}}}

\newcommand{\mr}[1]{\mathrm{#1}}

\newcommand{\cop}{\mr{COP}}

\usepackage{xspace}

\newcommand{\wh}{excess heat }
\newcommand{\Wh}{Excess heat }
\newcommand{\ssc}{self-scheduling\xspace}

\newcommand{\mpa}{market participation\xspace}

\newcommand{\prds}{producers\xspace}
\usepackage{xcolor}

\usepackage{pifont}
\usepackage{gensymb} 

\usepackage{siunitx}

\usepackage{amssymb}
\usepackage{amsfonts}

\usepackage{subcaption}

\usepackage{tikz}
\usetikzlibrary{decorations.text}
\usetikzlibrary{calc}
\usetikzlibrary{positioning}
\usetikzlibrary{decorations.markings}
\usetikzlibrary{decorations.shapes}
\usetikzlibrary{decorations.text}
\usetikzlibrary{decorations.fractals}
\usetikzlibrary{decorations.footprints}
\usetikzlibrary{spy}
\usetikzlibrary{shapes.arrows}
\usetikzlibrary{arrows,shapes,automata,backgrounds,petri,positioning}
\usetikzlibrary{arrows.meta}
\usetikzlibrary{decorations.pathmorphing}
\usetikzlibrary{shapes.arrows}
\usetikzlibrary{shadows}
\usetikzlibrary{calc}
\usetikzlibrary{spy}

\usepackage{nomencl}
\usepackage{etoolbox}
\usepackage{multicol}
\newlength{\nomitemorigsep}
\makenomenclature
\usepackage{etoolbox}
\renewcommand\nomgroup[1]{%
  \item[\bfseries
  \ifstrequal{#1}{I}{Super- and subscripts}{%
  \ifstrequal{#1}{S}{Sets}{%
  \ifstrequal{#1}{P}{Parameters}{%
  \ifstrequal{#1}{V}{Variables }{}}}}%
]}
\usepackage[font=small]{caption}

\usepackage[a4paper,top=2.2cm,bottom=2.2cm,left=2.2cm,right=2.2cm,marginparwidth=1.75cm]{geometry}
\usepackage{color}
\usepackage{hyperref}
\hypersetup{
     colorlinks = true,
     linkcolor = blue,
     anchorcolor = blue,
     citecolor = blue,
     filecolor = blue,
     urlcolor = blue
     }
     
\usepackage[cmex10]{amsmath}
\usepackage{amssymb}
\usepackage{amsmath}
\usepackage{amsthm}

\usepackage{float}
 
\usepackage{cite}
\usepackage{eurosym}
\usepackage{enumitem}
\makeatletter\DeclareRobustCommand\onedot{\futurelet\@let@token\@onedot}
\def\@onedot{\ifx\@let@token.\else.\null\fi\xspace}

\usepackage[final]{changes}

\begin{document}

\title{\huge {\bf Market Integration of Excess Heat}}

\author{\Large Linde Frölke$^*$, Ida-Marie Palm, Jalal Kazempour \\ \textcolor{white}{.} \\ \small  Technical University of Denmark \\ \small Department of Wind and Energy Systems \\ \small $^*$ Corresponding Author Email: \href{mailto:linfr@dtu.dk}{linfr@dtu.dk} }
 
\date{\today}
 
\maketitle

\maketitle


\begin{abstract}
\Wh will be an important heat source in future carbon-neutral district heating systems. A barrier to \wh integration is the lack of appropriate scheduling and pricing systems for these \prds, which generally have small capacity and limited flexibility. In this work, we formulate and analyze two methods for scheduling and pricing \wh \prds: \ssc and \mpa. In the former, a price signal is sent to \wh \prds, based on which they determine their optimal schedule. The latter approach allows \wh producers to participate in a market clearing. In a realistic case study of the Copenhagen district heating system, we investigate market outcomes for the two excess heat integration paradigms under increasing excess heat penetration.
An important conclusion is that in systems of high \wh penetration, simple price signal methods will not suffice, and more sophisticated price signals or coordinated dispatch become a necessity. 

\vspace{0.3cm}

\end{abstract}
\vspace{1cm}


\section{Introduction}
District heating is expected to play an important role in future carbon-neutral energy systems, especially in urban areas \cite{Lund2014FourthGDH}. 
District heating networks facilitate the decarbonization of heat generation, for instance by allowing for distribution of \textit{excess heat} to consumers. 
Examples of excess heat \prds include energy intensive industries such as metal and cement factories, sources in the service sector that produce heat as a by-product of their refrigeration systems \cite{Zulsdorf2018Analysis}, and data centers that produce heat from cooling their servers \cite{Wahlroos2018Future}.  
\deleted{It is estimated that the total excess heat potential in the European Union is about 300 TWh per year, around 10 TWh of which \added{is }located in Denmark.} 
In many cities, excess heat has the potential to cover a large share of total heat demand. 
\added{For example, \cite{Amer2019Modelling} finds that excess heat could cover over half of Greater Copenhagen's heat demand in 2050. 
Excess heat could also cover over 80\% of demand in several other large Danish district heating networks \cite{Buhler2017Industrial}.}
Current district heating systems typically rely on few large generators, such as Combined Heat and Power (CHP) plants and waste incinerators. Compared to these conventional sources, excess heat \prds are generally of smaller capacity and lower flexibility.
Integration of excess heat \prds would therefore result in a more distributed heating system, with a large number of small heat sources.
In such a system, coordination and scheduling of heat generation becomes more challenging. 

A major barrier to excess heat injection in district heating networks is the lack of suitable methods for \textit{scheduling} and \textit{pricing} excess heat.  
Most potential excess heat sources are untapped, even though it has been shown that their integration decreases both fuel usage and operational cost of the system \cite{Syri2015OpenDH}, and the integration of excess heat has proven feasible in simulation studies and in practice \cite{Brand2014Smart, Buffa2019Fifth}. It remains an open question how heat scheduling and pricing systems can be designed to optimally integrate excess heat producers.

\added{Most existing district heating systems do not have a liberalized market. In Greater Copenhagen, the daily heat dispatch is determined by Varmelast.dk, {which is a regulated heat market} that dispatches generators based on submitted price-quantity bids. 
While the scheduled quantities are determined by this market, the prices that could be derived from the market are not used. Instead, the price of heat is fixed in advance in contracts between suppliers of district heating and distributors/transmitters. Excess heat providers currently do not have the possibility to participate in this dispatch procedure. See e.g. \cite{Wang2019Investigation} for more details on Varmelast.dk.}

{
Only few existing works have studied excess heat producers in a market setting. 
In \cite{Syri2015OpenDH}, the impact of excess heat producers on the heat market in Espoo, Finland, is investigated. 
References \cite{Dominkovic2018Influence} and \cite{Doracic2021UtilizingExcess} study the potential effect of dynamic pricing on marginal cost of district heating systems, for a system including excess heat producers. In both works, it is assumed that the excess heat production profile is constant over each month or over the entire year, and that this profile is fully inflexible. 
Furthermore, the price bidding behaviour of CHPs is not modeled accurately, either disregarding electricity price dependence \cite{Doracic2021UtilizingExcess}, or disregarding the dependence on opportunity cost in the electricity market \cite{Dominkovic2018Influence}. 
Reference \cite{Liu2019MarginalCost} applies marginal-cost pricing to a case study in the Netherlands to assess whether different producers can recover their fixed costs from market revenues. Their market clearing consists of a combined unit commitment and economic dispatch. Excess heat from industrial processes is included, but its flexibility is not modeled. 
None of these works investigate the effect of increasing excess heat penetration. }

{In this work, we investigate whether price signals to be disseminated by the heat market operator can suffice for market integration of excess heat \prds.
To the best of our knowledge, this is the first paper in the literature that explores the integration of excess heat producers through price signals. 
Given a price signal, excess heat producers self-schedule their production. We evaluate this \textit{self-scheduling} model by comparison to an ideal benchmark, namely direct \textit{market participation} of excess heat \prds. This comparison is done by performing a realistic case study of the Copenhagen district heating system, which currently includes 13 CHP plants.}
We evaluate the success of the self-scheduling method by adding an increasing number of excess heat producers to the Copenhagen system. 
We aim to show the consequences of integrating cooling-based excess heat \prds under these two paradigms, including how the suboptimality of the \ssc model evolves under increased penetration of excess heat. 
{Our main finding is that price signals can be used as an alternative for market participation of excess heat producers, but their success depends highly on the quality of the signal, as well as the penetration of excess heat. As long as the installed excess heat capacity is sufficiently low, a simple price signal may suffice, but as excess heat penetration increases there may be significant downsides to this approach.}



The remainder of this article is structured as follows. In Section \ref{sec:market&self}, we discuss the self-scheduling and market participation models in more detail. We provide model formulations, and outline the bidding behavior of different market participants. The real case study of the Copenhagen district heating system is presented in Section \ref{sec:results}, including numerical results. In Section \ref{sec:conclusion} we conclude with further discussions regarding the implications of both methods and provide several recommendations for future work. 

\section{Market participation versus self-scheduling}\label{sec:market&self}
One natural way of integrating excess heat producers in heat markets, is by direct market participation. For example in the district heating system of Copenhagen, a marginal-cost based heat market is operated by Varmelast\footnote{www.varmelast.dk/}. 
This paradigm will further be referred to as \textit{\mpa}. 
Another option is for heat market operators to publish a time-varying price signal for excess heat \prds, which optimally self-schedule their heat generation accordingly, and share the resulting schedule with the market operator. We further refer to this paradigm as \textit{\ssc}. In the Open District Heating system of Stockholm, small excess heat producers are successfully integrated in this way using an ambient temperature-dependent price signal\footnote{\url{www.opendistrictheating.com}}. 

We first discuss the advantages and benefits of each model in Section \ref{sec:2par}, and then present the model formulations in Sections \ref{sec:mpa} and \ref{sec:ssc}. Finally, Section \ref{sec:bidding} outlines CHP and excess heat provider bidding behaviour. 

\subsection{Comparison of market integration and self-scheduling}\label{sec:2par}
We compare the two paradigms in Table \ref{tab:compare}.
First of all, they differ in the formation of the price received by excess heat participants. Under the \mpa scheme, the market-clearing price follows from the bids submitted by market participants (including excess heat \prds), while the price for excess heat is set exogenously by the market operator in the \ssc case. 
As also indicated in Table \ref{tab:compare}, market participation would provide incentives to optimally schedule excess heat production, as it minimizes total generation cost. 
If the price signal for \ssc is designed perfectly, the resulting schedules may be the same as in a market setting. Otherwise, the schedule resulting from price signals will be suboptimal. This suboptimality may increase with the penetration of excess heat producers. 

Although \mpa of excess heat producers would be optimal from a cost minimization perspective, it has some drawbacks in practice for a heat system with many small (excess) heat \prds. For small excess heat \prds, it may be difficult to decide on market bids, and they may therefore prefer to receive a price signal. For the market operator, the \mpa of many small excess heat \prds poses a communicational challenge: the operator receives bids from many participants, clears a more complex market, and then needs to send the individual schedules to each small market participant. 
Therefore, the \ssc paradigm may be preferred in practice, as it is a rather simple and computationally cheap way of scheduling and pricing excess heat, with lower IT requirements. 
We consider the market participation scheme as \textit{ideal benchmark}, and explore the success of self-scheduling scheme in comparison to this ideal benchmark. 
\added{Both the market participation and the self-scheduling model are Linear Programs, in which no binary variables are used. }

\subsection{Model formulation for \mpa}\label{sec:mpa}
\begin{table}[t]
    \centering
    \begin{tabular}{c|c|c}
         & Market Participation & Self-scheduling \\
         \hline 
         Price formation & Endogenous & Exogenous \\
         Optimal scheduling & \checkmark  & \checkmark $\,$ / $\times$  \\
         Relevance for small producers & $\times$ & \checkmark \\
    \added{Problem type} & \added{Linear program} & \added{Linear program}
    \end{tabular}
    \caption{Comparison of the two paradigms for scheduling and pricing excess heat}
    \label{tab:compare}
\end{table}

In the market participation scheme, excess heat producers participate in a market clearing. 
We formulate a heat market clearing without network constraints as a linear optimization problem. The market clearing results in a uniform market price and scheduled quantities for all participants, including excess heat producers. 
All market participants submit price-quantity bids. 
The market then dispatches generators to minimize total generation cost, i.e., according to the merit order. 
\added{Following EU electricity market design, we do not consider unit commitment constraints. This implies that unit commitment constraints should be internalized \added{into} the bids. }
\added{We also choose not to {enforce} network constraints in the optimal dispatch, because the current optimal dispatch mechanism in Copenhagen does not include such constraints either. Instead, hydraulic conditions are checked after clearing the market \cite{Wang2019Investigation}.} 

The objective of the market clearing is to minimize \added{total heat} generation cost, given by the function
\begin{align}
    f(
    \Gamma\up{mp}) = \sum_{t\in\mathcal{T}} \biggl( \sum_{e \in \mathcal{E}} c_{et}(G_{et}\up{H} - W_{et} )  + \sum_{i \in \mathcal{G}} c_{it}G\up{H}_{it}  
    + c\up{U} U_t \biggr) \,,
\end{align}
where $\Gamma\up{mp}$ is the set of optimization variables for the market participation model, $c_{it}$ is the bid price of CHP $i$ at time $t$, $c_{et}$ is the bid price of excess heat producer $e$, $G\up{H}_{it}$ is the generated heat by a CHP, and  $c\up{U}$ represents the (constant) penalty cost per unit of unsupplied load $U_t$. The excess heat production $G_{et}\up{H}$ may in some cases exceed the load, so that an amount of $W_{et}$ will have to be wasted, i.e., vented to the outside air.
The wasted excess heat must be non-negative and cannot exceed the produced excess heat:
\begin{align}
    G_{et} \geq W_{et} \geq 0 && \forall e, t \,.
\end{align}

A power balance must hold between the predicted heat load $\hat{L}\up{H}_t$ and the scheduled generation. If the load cannot be supplied due to insufficient installed capacity, the load can be curtailed by an amount $U_t$ of unsupplied load:
\begin{align}\label{eq:powerbalance}
    \sum_{e\in \mathcal{E}} G_{et}\up{H} - W_{et} + \sum_{i\in \mathcal{I}} G_{it}\up{H} = \hat{L}\up{H}_t - U_t \quad : \lambda\up{H}_t && \forall t \,.
\end{align}
The uniform market price $\lambda\up{H}_t$ is given by the dual variable corresponding to constraint \eqref{eq:powerbalance}, which is equal to the marginal price bid of the most expensive scheduled generator. 

\begin{figure}[t]
    \centering
\tikzstyle{BOXY_S} = [rectangle, rounded corners = 5, minimum width=20, minimum height=15,text centered, draw=black, fill=white,line width=0.3mm,font=\footnotesize]
\tikzstyle{BOXY_R} = [rectangle, rounded corners = 5, minimum width=20, minimum height=20,text centered, draw=black, fill=white,line width=0.3mm,font=\footnotesize]
\tikzstyle{BOXY_HS} = [rectangle, minimum width=0, minimum height=20,text centered, fill=white,line width=0.3mm,font=\footnotesize]
\begin{tikzpicture}[node distance=60]
    \node [align=center ] (box1S) [BOXY_S] {Bid submission \\  $\forall e, i,t$};
    \node [align=center] (HS1) [BOXY_HS, left = 0.4cm of box1S, xshift=0cm] {$\lambda\up{E}_t$};
    \draw[->,>=stealth,line width=0.3mm, color=black] (HS1) -- (box1S);
    
    
    \node [align=center ] (box2S) [BOXY_S, right = 1.5 cm of box1S] {MO clears \\ market};
    \node [align=center] (HS2) [BOXY_HS, right = 0.4cm of box2S] {$G\up{H}_{it}$, $G\up{H}_{et}$, \\ $\lambda\up{H}_t$};
    \draw[<-,>=stealth,line width=0.3mm, color=black] (HS2) -- (box2S);
    \path (box1S) edge[->,>=stealth,line width=0.3mm] node[yshift=2.5mm, text=black]{$c_{it}, \mathcal{F}_{it}$} (box2S);
    \path (box1S) edge[->,>=stealth,line width=0.3mm] node[yshift=-2.5mm, text=black]{$c_{et}, \mathcal{F}_{et}$} (box2S);
\end{tikzpicture}
    \caption{Overview diagram of market participation model. Market Operator is abbreviated to MO. }
    \label{fig:tikz_mpa}
\end{figure}
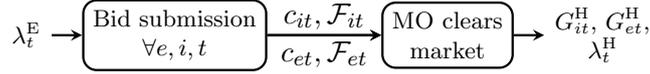

The market participants' bid includes a description of the feasible region $\mathcal{F}$ of their heat generation. These feasible regions are respected in the market clearing:
\begin{align}
    G\up{H}_{it} &\in \mathcal{F}_{it} && \forall i,t \\
    G\up{H}_{et} &\in \mathcal{F}_{et} && \forall e,t \,. 
\end{align}
We will define these feasible regions for CHPs and excess heat producers in Section \ref{sec:bidding}, and in more detail in Appendices \ref{ap:CHPmodel} and \ref{ap:eh_model}.
The set of optimization variables in the market-clearing optimization problem is given by $\Gamma\up{mp} = \{G_{it}\up{H} ,  G\up{H}_{et}, W_{et}\up{H} ,  U_{t}\up{H}\}$.
Fig. \ref{fig:tikz_mpa} shows a graphical overview of the market participation model. \added{The forecasted electricity price $\lambda_t\up{E}$ is an input to the marginal cost model of both CHPs and excess heat producers, which will be discussed in Section \ref{sec:bidding}.} 

\subsection{Model formulation for self-scheduling}\label{sec:ssc}
In the self-scheduling model, the market operator broadcasts a price signal $\mu_t\up{H}$ for each market period $t$ to all excess heat producers, who self-schedule their production accordingly. The resulting schedule is submitted to the market operator, who uses the total excess heat production as a fixed input to the market clearing with conventional generators only. This market clearing is as described previously in Section \ref{sec:mpa}, except that $G\up{H}_{et}$ is now a parameter instead of a variable for all $e,t$. 
As a result, the total self-scheduled heat generation by the \wh \prds is prioritized in the heat market, and the CHPs may supply any remaining unsupplied load. If the self-scheduled excess heat exceeds the heat load at certain hours, some of the excess heat is wasted, i.e., vented to the air. 

We assume an ambient-temperature-dependent price signal for excess heat producers, inspired by the Stockholm Open District Heating pricing system. In particular, the received price decreases with the ambient temperature, as heat demand often decreases with ambient temperature too. 
Under the self-scheduling scheme, the excess heat producers are paid as in the price signal for each unit generated, regardless of whether (part of) the produced excess heat exceeds the supplied load and needs to be wasted. \added{Clearly, this is an undesirable effect of the self-scheduling paradigm.} The CHPs are still paid \added{at} the \added{uniform} marginal price resulting from the market clearing. 


During self-scheduling, \wh \prds aim to minimize total cost, given by the difference between costs for electricity used by the heat pump $L\up{E}_{et}$ bought at the (forecasted) electricity spot price $\lambda\up{E}$, and the income from selling the generated heat $G\up{H}_{et}$:
\begin{align}
    C\up{H}_{e} = \sum_{t\in\mathcal{T}} \left ( \lambda\up{E}_t L\up{E}_{et} - \mu\up{H}_t  G\up{H}_{et}  \right ) \,.
\end{align}
It is assumed that all excess heat producers and CHPs use the same forecasted electricity prices $\lambda_t\up{E}$. 
The \wh \prds must schedule an amount of heat that respects their physical constraints:
\begin{align}
    G\up{H}_{et} \in \mathcal{F}_{et} && \forall e,t \,.
\end{align}
The feasible region $\mathcal{F}_{et}$ will be defined in the next Section \ref{sec:bidding}, and more details can be found in Appendix \ref{ap:eh_model}.
Fig. \ref{fig:tikz_ssc} shows a graphical overview of the self-scheduling model.

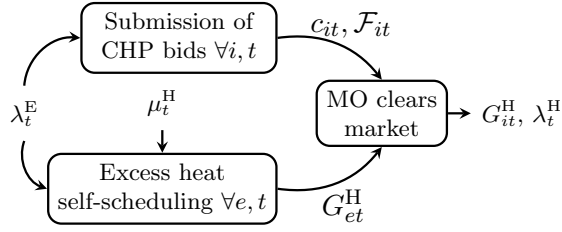
\begin{figure}[t]
    \centering
\tikzstyle{BOXY_S} = [rectangle, rounded corners = 5, minimum width=20, minimum height=15,text centered, draw=black, fill=white,line width=0.3mm,font=\footnotesize]
\tikzstyle{BOXY_R} = [rectangle, rounded corners = 5, minimum width=20, minimum height=20,text centered, draw=black, fill=white,line width=0.3mm,font=\footnotesize]
\tikzstyle{BOXY_HS} = [rectangle, minimum width=0, minimum height=20,text centered, fill=white,line width=0.3mm,font=\footnotesize]
\begin{tikzpicture}[node distance=60]
    \node [align=center ] (mo) [BOXY_S] {MO clears \\ market};
    \node [align=center] (outputs) [BOXY_HS, right = 0.3cm of mo] {$G\up{H}_{it}$, $\lambda\up{H}_t$};
    \draw[<-,>=stealth,line width=0.3mm, color=black] (outputs) -- (mo);
    \node [align=center ] (chps) [BOXY_S, left = 0.5 cm of mo, yshift = 1cm, minimum width=73] {Submission of \\ CHP bids $\forall i,t$};
    \path (chps.east) edge[->,>=stealth,line width=0.3mm, bend left=30] node[yshift=2.5mm, xshift=2mm,  text=black]{$c_{it}, \mathcal{F}_{it}$} (mo.north);

    \node [align=center ] (selfsched) [BOXY_S, left = 0.5 cm of mo, yshift = -1cm, minimum width = 73] {Excess heat \\ self-scheduling $\forall e,t $};
    \path (selfsched.east) edge[->,>=stealth,line width=0.3mm, bend right=30] node[yshift=-3mm, xshift=1mm,text=black]{$G\up{H}_{et}$} (mo.south);
    
    \node [align=center] (lambda) [BOXY_HS, left = 3.5 cm of mo, xshift=0cm] {$\lambda\up{E}_t$};
    \node [align=center] (mu) [BOXY_HS, above = 0.3 cm of selfsched, xshift=0cm] {$\mu\up{H}_t$};
    \draw[->,>=stealth,line width=0.3mm, color=black] (mu.south) -- (selfsched.north);
    \path (lambda.north) edge[->,>=stealth, line width= 0.3mm, bend left=45, draw=black] (chps.west);
    \path (lambda.south) edge[->,>=stealth, line width= 0.3mm, bend right=45, draw=black] (selfsched.west);
    
\end{tikzpicture}
    \caption{Overview diagram of self-scheduling model and the following market clearing. Market Operator is abbreviated to MO. }
    \label{fig:tikz_ssc}
\end{figure}

\subsection{Excess heat and CHP models}\label{sec:bidding}
To simulate the two paradigms, the bidding behavior of different market participants needs to be modeled. 
In this work, we consider CHPs and cooling-based excess heat producers only. The derivation of their bidding behavior and feasible regions is presented in more detail in Appendices \ref{ap:CHPmodel} and \ref{ap:eh_model}. We assume that the heat market is cleared daily before the electricity market is cleared, as is currently the case in Copenhagen. 
The resulting feasible region for CHP $i$ at time $t$ is given by
\begin{align}\label{eq:chp_feasible}
    \mathcal{F}_{it} = \left \{ G\up{H}_{it} \quad \middle | \quad 0 \leq  G\up{H}_{it} \leq \min \left (\overline{G}\up{H}_i, \frac{\overline{F}_i}{\rho\up{H}_i + r_i \, \rho\up{E}_i} \right  ) \right \}  \,, 
\end{align}
where $\overline{G}\up{H}_i$ is the maximum heat generation, $\rho\up{E}_{i}$ and $\rho\up{H}_{i}$ are the fuel efficiency for electricity and heat, respectively, $r_i$ is the minimum power-to-heat ratio, and $\overline{F}_i$ is the maximum fuel consumption. Note that these parameters are here considered time-invariant, but this could easily be adapted. 

Reference \cite{Mitridati2020Heat} derives the optimal heat bid $c\up{H}_{it}$ for CHPs in a sequential heat and electricity market setting. The marginal cost of heat depends on the (forecasted) electricity price $\lambda\up{E}$ \added{and (constant) fuel price $\alpha_i$ }as follows:
\begin{align}\label{eq:chp_cost}
    c\up{H}_{it} = 
    \begin{cases}
        \alpha_i \, (\rho\up{E}_i \,r_i + \rho\up{H}_i) - \lambda\up{E}_t \, r_i & \text{if } \lambda\up{E}_t \leq \alpha_i \rho_i\up{E} \\
        \lambda\up{E}_t \, \frac{\rho\up{H}_i}{\rho\up{E}_i} & \text{if } \lambda\up{E}_t > \alpha_i \rho\up{E}_i \,.
    \end{cases}
\end{align}
This bidding function assumes \added{that in case the forecasted electricity price is lower than {or equal to} a certain threshold, CPHs bid their fuel cost minus the income from electricity sale, as represented by the first case in \eqref{eq:chp_cost}. }\replaced{Otherwise,}{that in case} if the forecasted electricity price is relatively high, CHPs bid the lost opportunity cost from selling heat instead of electricity\added{, as in the second case in \eqref{eq:chp_cost}}.

The feasible region for excess heat producers is derived in Appendix \ref{ap:eh_model}. 
The excess heat \prds' flexibility in heat production is represented by modeling cooling cabinet heat dynamics as a set of linear constraints as given in Appendix \ref{ap:eh_model} through  Eqs. \eqref{eq:ex1}-\eqref{eq:exend}. 
We define the feasible region for excess heat production as the set of production setpoints that satisfy the aforementioned constraints:
\begin{align}\label{eq:eh_feasible}
    \mathcal{F}_{et} = \left \{G\up{H}_{et} \quad \middle | \quad  \eqref{eq:ex1}-\eqref{eq:exend} \right \} \,.
\end{align}

The price bid of excess heat producers in the market participation model is assumed to be at zero\footnote{Bidding at zero is reasonable if any sold excess heat is considered extra income for these producers. However, the electricity consumption cost of a given excess heat production profile may be greater than the electricity cost of the production profile that minimizes these costs. One may therefore choose to define the price bid of a certain production profile as the difference between the electricity cost of the given profile and the minimum electricity cost this producer could obtain if it would minimize electricity cost only. }, i.e.,
\begin{align}
    c_{et} = 0 && \forall e,t \,.
\end{align}


\section{Copenhagen case study}\label{sec:results}
We analyze the application of the market participation and self-scheduling models in the district heating system of Copenhagen, Denmark. 

\subsection{Case study description}
The district heating system of Copenhagen consists of 13 CHPs. In our case study, we vary the level of excess heat capacity added to this system from 0 to 2100 MW. 
{The excess heat is assumed to be produced as a by-product of cooling. In particular, we assume excess heat producers cool refrigeration cabinets using a local heat pump. }
We simulate in an hourly time resolution for a full year. 
In our online appendix, a detailed description of all used parameters can be found, as well as the code used to generate our results\footnote{\url{www.github.com/linde-fr/excess-heat-in-market}}.

Several time series are needed as inputs for the market participation and/or self-scheduling models. 
The forecasted ambient temperature is an input used to model the Coefficient Of Performance (COP) of the heat pumps, and also to determine the price of waste heat in the self-scheduling model. We use hourly temperature measurements from 2019 from the Danish Meteorological Institute \cite{DMI_2021}. For the month March we use measurements from 2020, due to many missing measurements for March 2019. 
As the electricity price forecast, we use Nord Pool historical electricity prices for DK2 \cite{NordPoolData}.
For the forecasted heat load, we use the hourly heat load in the entire Copenhagen district heating area for 2019, provided by Varmelast. 

For the self-scheduling model, the excess heat price signal needs to be given.
The self-scheduling pricing signal used here is inspired by Stockholm's Open District Heating Spot Prima price. We approximate their ambient-temperature dependent price function using an exponential regression on data available from their website. The resulting price signal $\mu\up{H}_t(\cdot)$ is defined as follows:
\begin{align}
     \mu\up{H}_t(T\up{A}_t) = 
     \begin{cases}
        380 \cdot 0.92^{T\up{A}_t} & \text{for }T\up{A}_t < 17.5 \celsius \\
        0 & \text{for } T\up{A}_t \geq 17.5 \celsius \,.
     \end{cases}
\end{align}
Note that the price is decreasing with the ambient temperature, as the base of the exponent $0.92$ is non-negative and below 1.

We further require input parameters for CHP and excess heat producer models. 
For the CHPs, most input parameters have been obtained from \cite{Ommen2013Heat}. The minimum power-to-heat ratio was not given there, and therefore a default value of $r=0.45$ has been taken from \cite{EurostatData2021}.

The excess heat producers are assumed to have the same input parameters. The heat dynamics parameters are $A= 0.1$ and $B = \frac{1}{21}$. The temperature in the cooled room has to be within 2-8 $\celsius$, while the average temperature every 6 hours has to be within 4-5 $\celsius$.
The indoor temperature at the excess heat \prds is assumed to be constant at 25 $\celsius$.
The heat pumps' maximum generation capacity is $\overline{G}\up{H} = 30$ \si{\kilo \watt}. Heat pump ramping limits are set to $0.25$ of its maximum generation capacity. 
It is assumed that the COP of the excess heat producers' heat pumps varies with the ambient temperature. The approximate ambient temperature dependence of the COP was obtained using a more detailed heat pump simulation model, under several simplifying assumptions, including a linear dependence of supply and return temperatures on ambient temperature. 

To validate our CHP bidding model, we have compared our resulting heat market prices to Varmelast heat market prices for a given electricity price signal. The results showed satisfactory correspondence between our and Varmelast's heat prices.\footnote{Exact results cannot be shared here due to confidentiality of Varmelast pricing data.}

\subsection{Results}
\added{The total heat load in 2019 in Copenhagen was around $8.3 \si{\tera \watt\hour}$ or $30 \si{\peta \joule}$. For the different maximum capacity levels of the excess heat producers of 300, 1200, and 2100 $\si{\mega\watt}$ participating in the heat market, we find that the excess heat providers are scheduled for $1.8$, $5.8$, and $8.0 \si{\tera\watt\hour}$, respectively. Considering that \cite{Amer2019Modelling} reports that excess heat could cover over $50\%$ of the Copenhagen heat load, the 1200MW case could be realistic for the Copenhagen system in 2050.}

\subsubsection*{Suboptimality}
\begin{figure}
    \begin{subfigure}[t]{0.48\textwidth}
        \includegraphics[width=\textwidth]{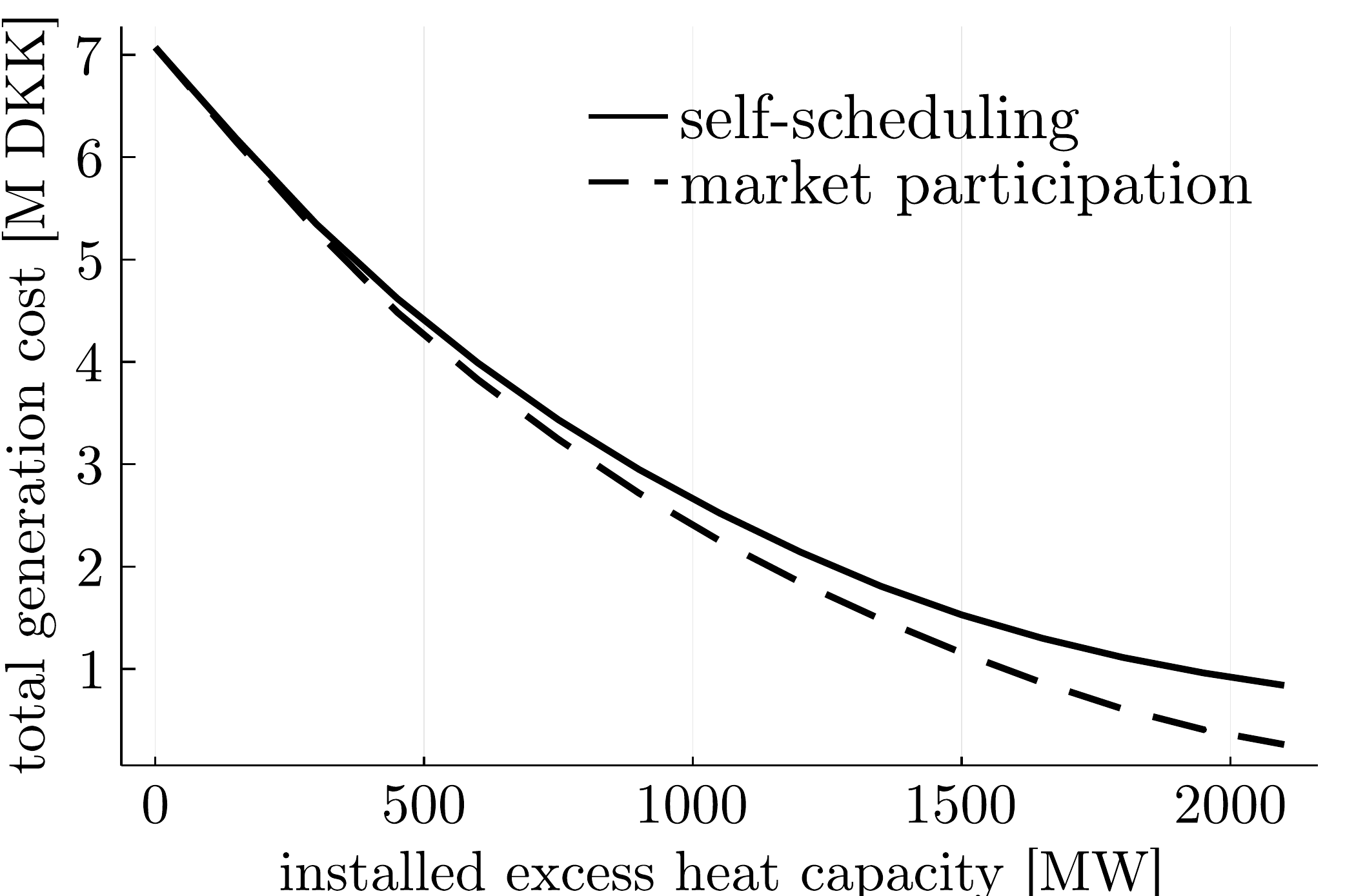}
        \caption{Total generation cost}
        \label{fig:totgen}
    \end{subfigure}
        \begin{subfigure}[t]{0.48\textwidth}
        \includegraphics[width=\textwidth]{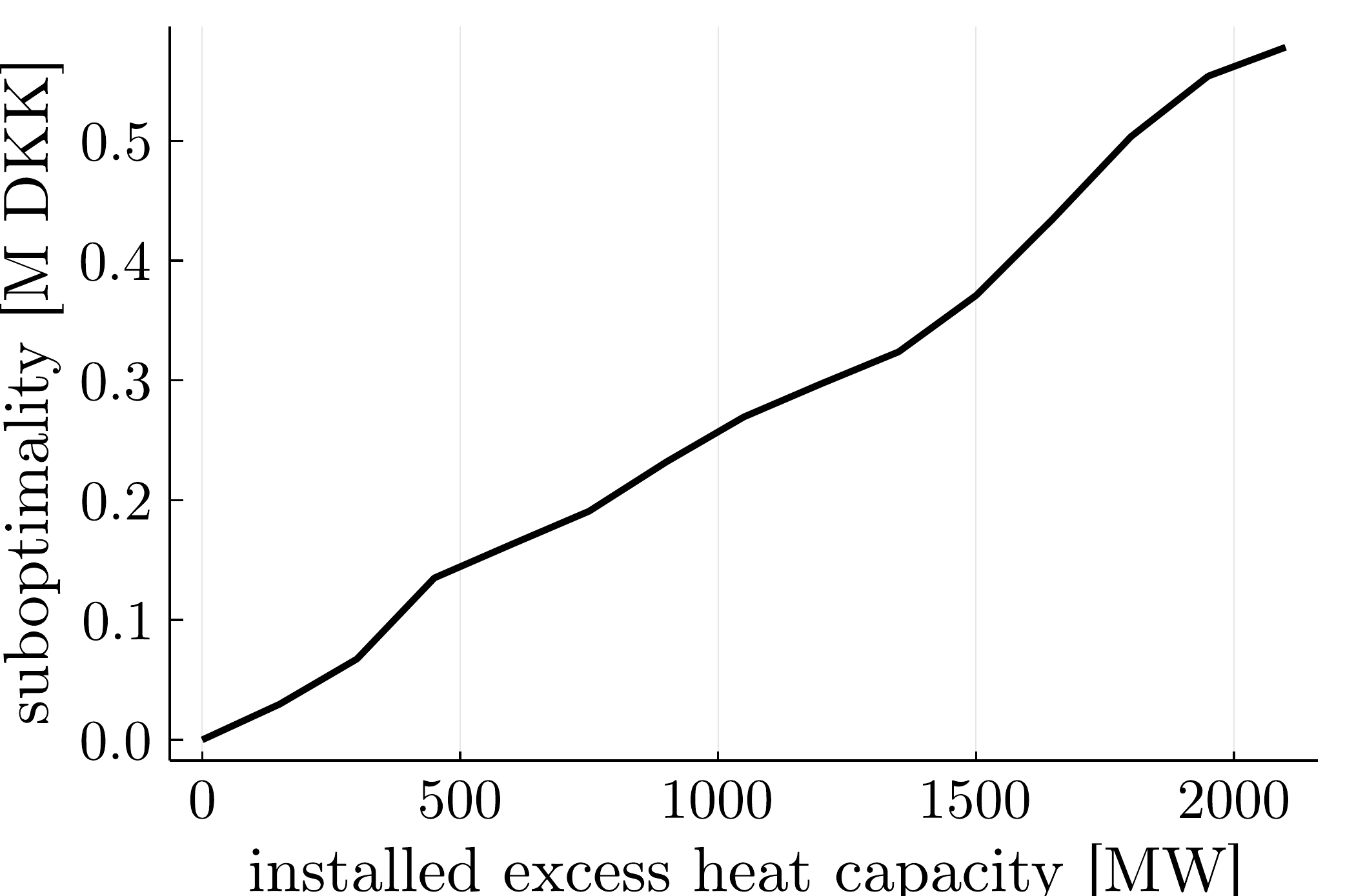}
        \caption{Difference in generation cost}
        \label{fig:diffgen}
    \end{subfigure}   
    \caption{Full year suboptimality of self-scheduling compared to market integration in terms of total CHP generation cost. \added{Note that this does \textit{not} include excess heat generation cost.}}
    \label{fig:suboptimality}
\end{figure}

When the excess heat \prds are \ssc and the market for CHPs is cleared afterwards, the total generation cost for CHPs will be greater than \added{it is} in the case the market is cleared with excess heat \prds integrated. 
Here we consider suboptimality in terms of the total CHP generation cost, where we treat this total cost similarly as in the objective function of the market clearing. That is, generation cost of each scheduled CHP is computed as its scheduled quantity multiplied by its price bid, and the total generation cost is obtained by summing over all CHPs. 
In Fig. \ref{fig:suboptimality}, we visualize how the total suboptimality over a full year depends on the penetration of excess heat \prds. In the left-hand Fig. \ref{fig:totgen}, we observe that both schemes experience a steep decrease of the total generation cost with the installation of the first 500 MW of excess heat capacity, but this decrease flattens out afterwards. 
Fig. \ref{fig:diffgen} shows that the absolute suboptimality grows almost linearly with the installed excess heat capacity. 
The curve is slightly steeper when the latest installed excess heat is replacing a CHP that is expensive compared to the CHPs that bid a price just below it. 

\begin{figure}
    \begin{subfigure}[t]{0.48\textwidth}
        \includegraphics[width=\textwidth]{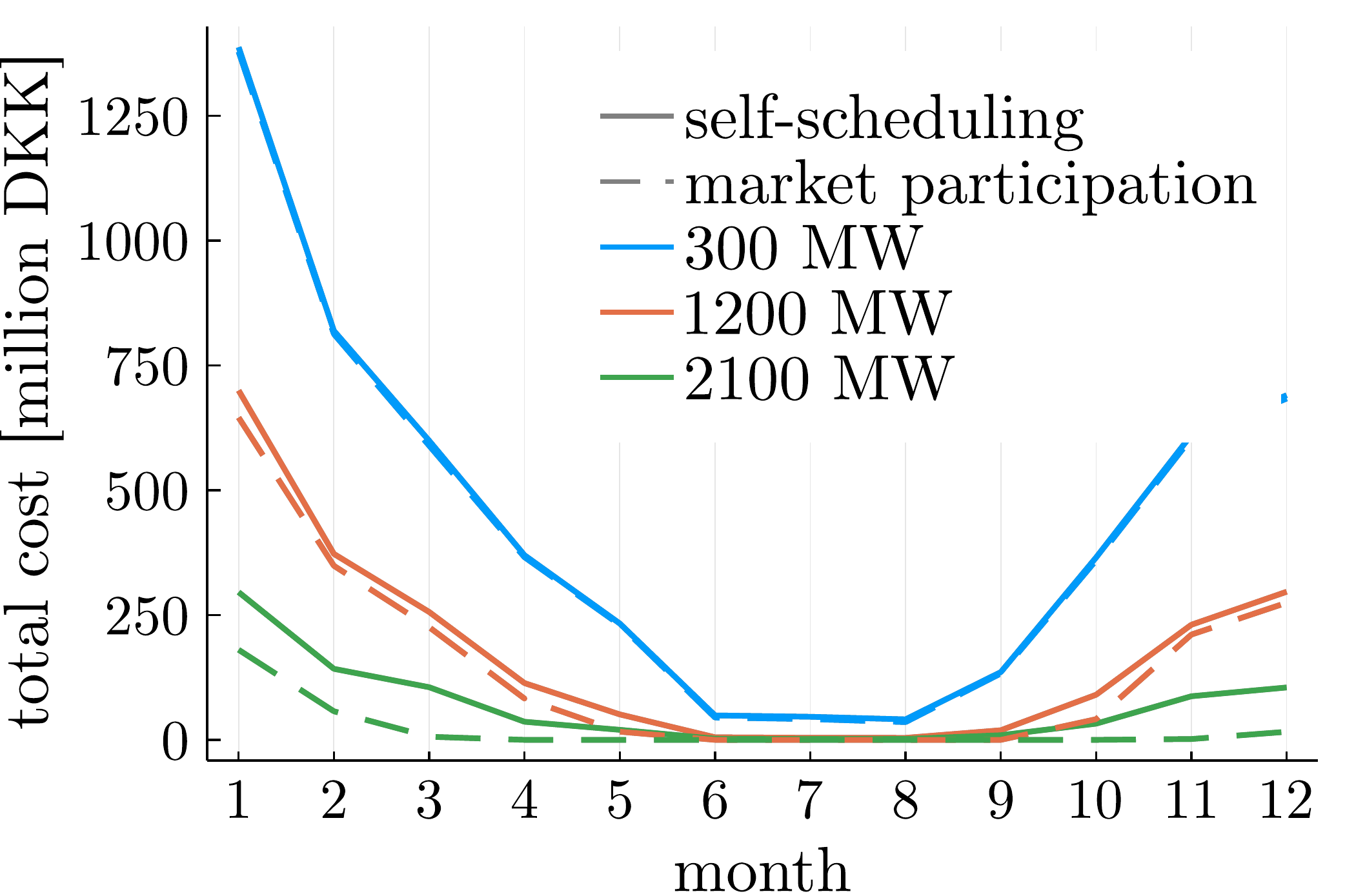}
        \caption{Total generation cost}
        \label{fig:totgen_m}
    \end{subfigure}
        \begin{subfigure}[t]{0.48\textwidth}
        \includegraphics[width=\textwidth]{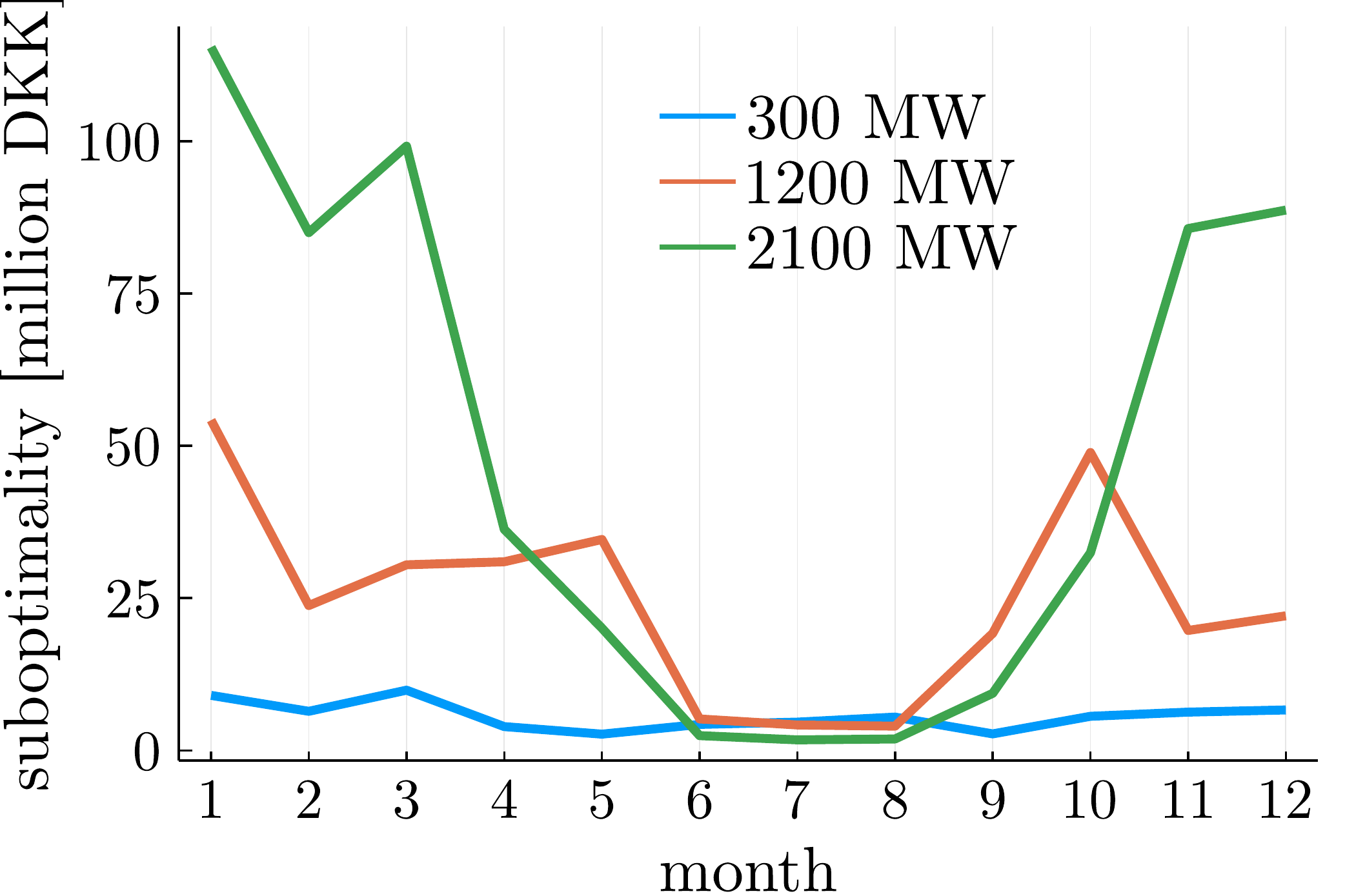}
        \caption{Difference in generation cost}
        \label{fig:diffgen_m}
    \end{subfigure}   
    \caption{Monthly suboptimality of self-scheduling compared to market integration in terms of total CHP generation cost.}
    \label{fig:suboptimality_m}
\end{figure}

Next, we investigate how this suboptimality in total CHP generation cost is distributed over the year, by zooming in on three levels of excess heat penetration in Fig. \ref{fig:suboptimality_m}. The left-hand Fig. \ref{fig:totgen_m} shows that the generation cost is unequally distributed over the year, as monthly heat load varies significantly over the year. As seen in the right Fig. \ref{fig:diffgen_m}, the suboptimality is quite equally distributed over the year for a low capacity of excess heat at 300 MW (blue). For higher excess heat penetration, the suboptimality is increasingly shifted to the colder months. 
In the warmest summer months, i.e., from June to August, the suboptimality of the 300 MW case (blue) is relatively high compared to the 1200 MW and 2100 MW cases (orange and green). The reason for this is that from a certain level of excess heat capacity, the overcapacity in summer is very high, so that the total demand is (almost) always completely supplied by excess heat, both under \ssc and market participation. Therefore, suboptimality will be low in summer for a higher penetration of excess heat. 

\subsubsection*{Scheduled and wasted excess heat}
\begin{figure}
    \begin{subfigure}[t]{0.48\textwidth}
        \includegraphics[width=\textwidth]{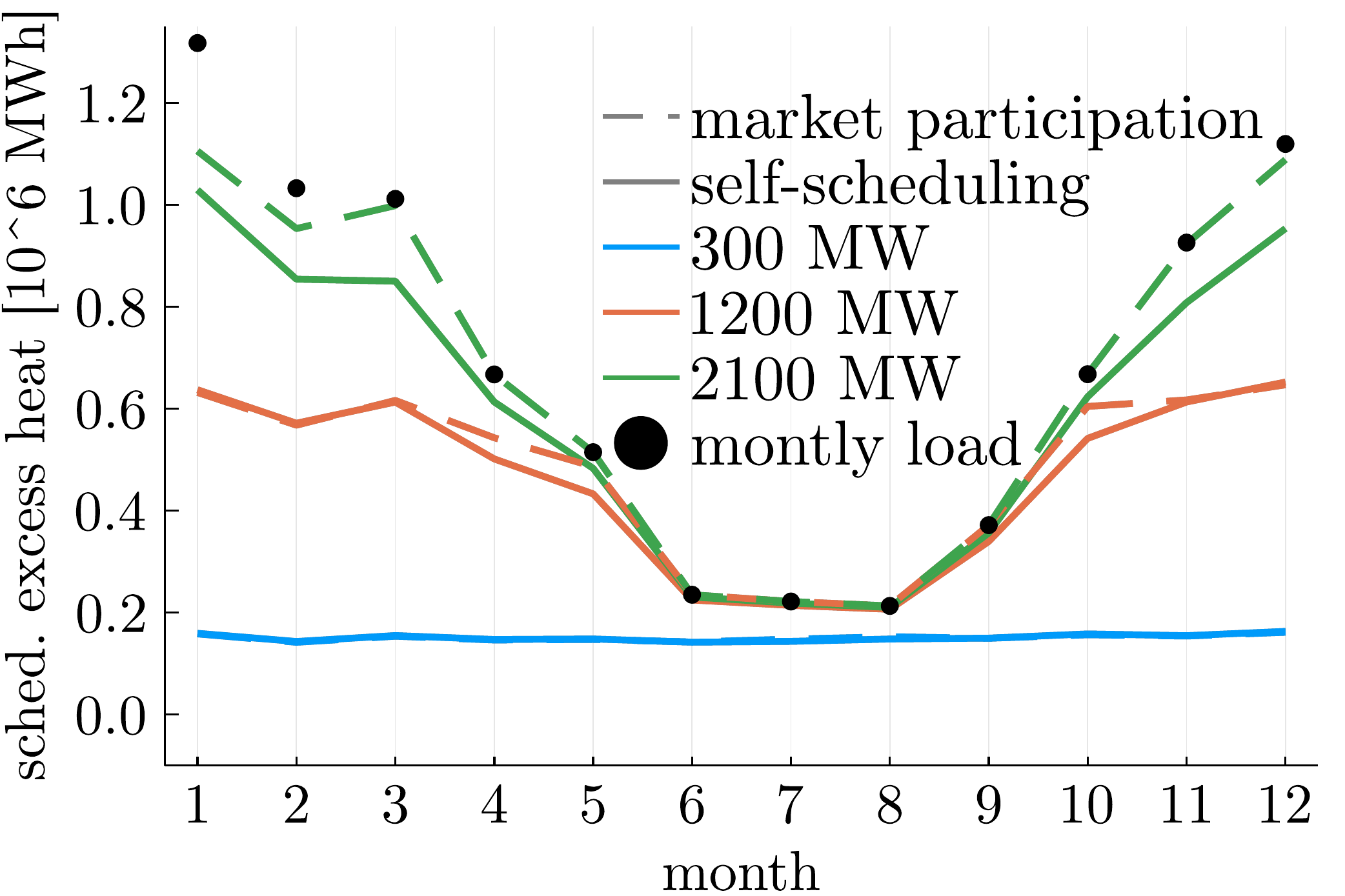}
        \caption{Scheduled volume}
        \label{fig:schedvol}
    \end{subfigure}
        \begin{subfigure}[t]{0.48\textwidth}
        \includegraphics[width=\textwidth]{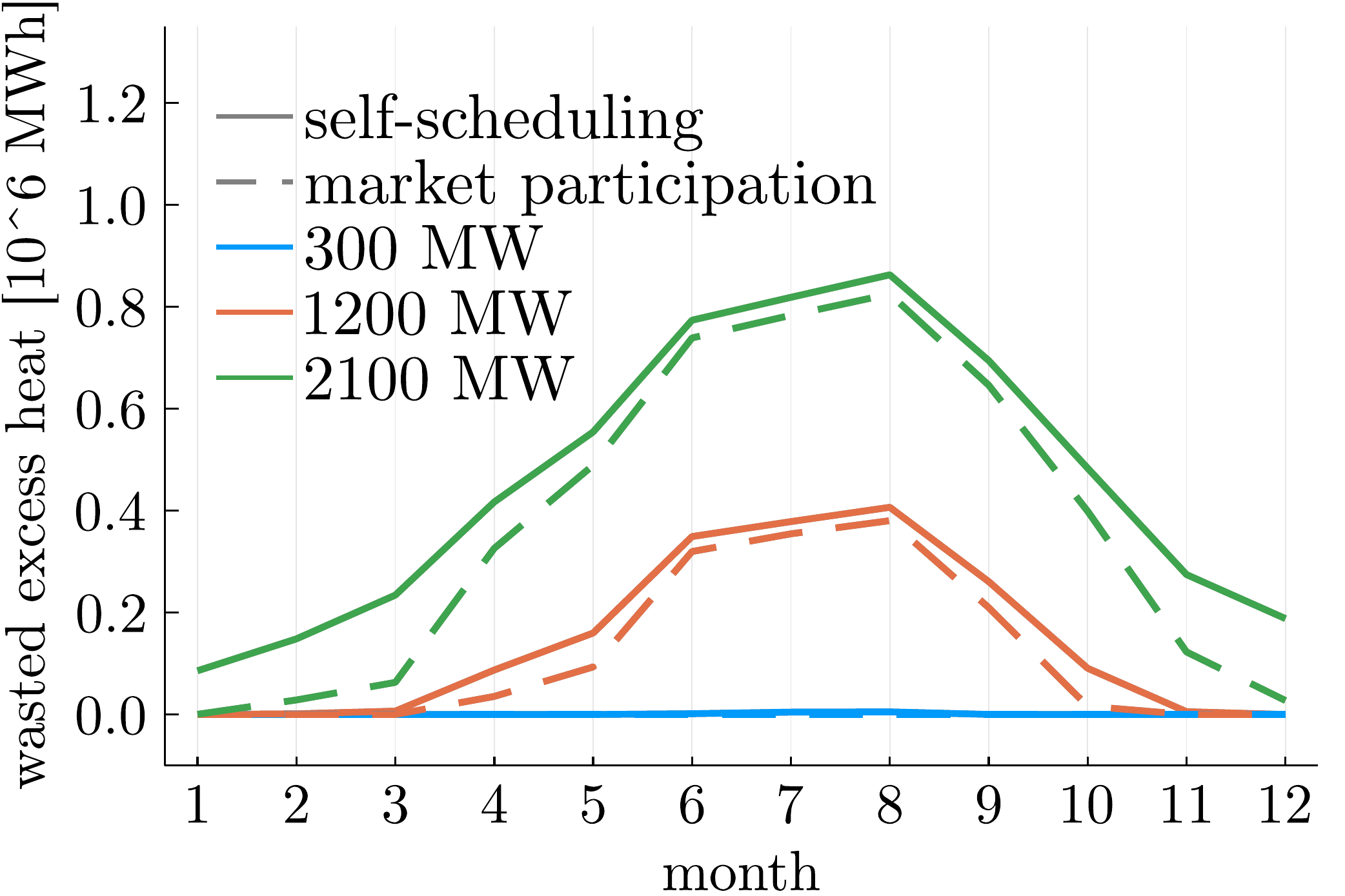}
        \caption{Wasted volume}
        \label{fig:wastedvol}
    \end{subfigure}   
    \caption{Monthly values for excess heat scheduled volume and wasted volume for self-scheduling compared to market integration}
    \label{fig:schedvol&wastedvol}
\end{figure}

The scheduled excess heat varies over the year. There can be a difference between \textit{generated} and \textit{scheduled} excess heat, as some excess heat may have to be wasted in case it exceeds the heat load. We compare the monthly scheduled excess heat for the \ssc and market participation under different levels of excess heat penetration in Fig. \ref{fig:schedvol}, and do the same for monthly wasted excess heat in Fig. \ref{fig:wastedvol}. Under the market participation model, a capacity of 2100 MW (green) is enough to supply the load fully in all but the coldest months, i.e., December to March. This is not the case for the \ssc model, which is due to a greater mismatch between supplied excess heat and heat load. This manifests itself in the consistently greater amount of wasted heat for the \ssc model. 
In general, differences in the \ssc and \mpa model in total scheduled excess heat are greatest in months where excess heat is the marginal supplier in some hours, but not in all hours. With increasing excess heat capacity, the months where this is the case shift more towards the winter months with greater heat load. 
Note that the scheduled volume in the summer months is almost equal for the 1200 MW (orange) and 2100 MW (green) cases, which is the reason that suboptimality in summer months is similar for these cases, as we observed previously in Fig. \ref{fig:diffgen_m}.

Finally, we highlight that the total wasted excess heat increases steadily with increasing excess heat penetration, for both scheduling paradigms. This is due to the limited flexibility of these \prds, combined with the fact that these producers have a minimum heat generation that may exceed the load. To decrease the wasted excess heat and supply a higher share of the load, it would be beneficial to install heat storage as the penetration of excess heat increases. 

\subsubsection*{Market prices}
\begin{figure}
    \centering
    \includegraphics[width=0.4\textwidth]{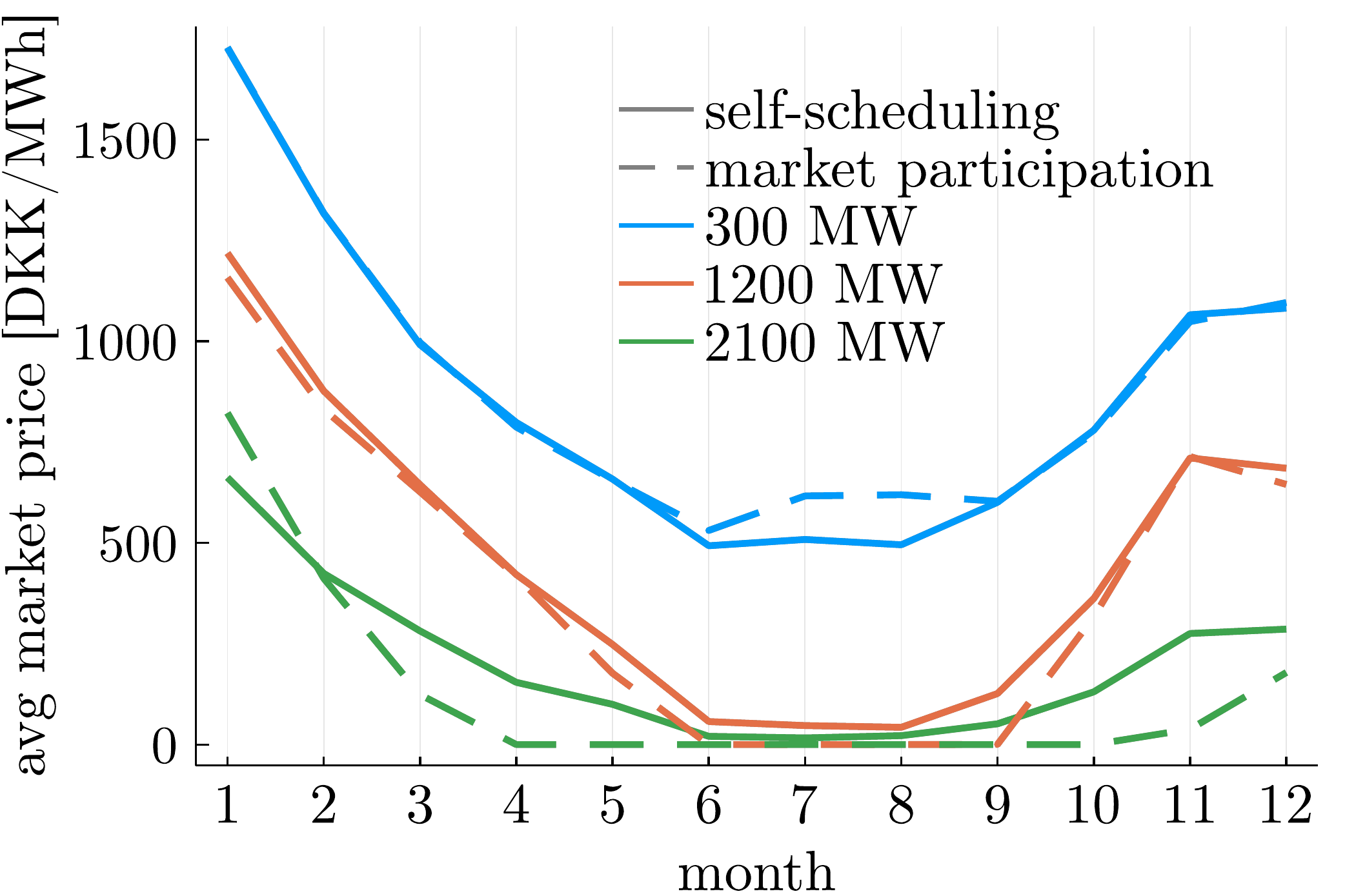}
    \caption{Average market prices per month for \ssc compared to \mpa, under various levels of excess heat penetration. }
    \label{fig:markprice}
\end{figure}

We compare monthly average market prices resulting from the market clearing in both the \ssc and \mpa models in Fig. \ref{fig:markprice}. Recall that we also clear the market in the \ssc model, but in this case the excess heat schedule is a fixed input to the market. Under a low excess heat penetration of 300 MW (blue), the average market price only differs for the summer months. Counter-intuitively, the average market price for the \mpa model is  higher in this case than that in the \ssc model. The explanation of this is that the \mpa model schedules excess heat when it can reduce total generation cost as much as possible, which means that it will spread out the excess heat schedule to avoid scheduling of more expensive generators. The effect of this spread can be that on average, the marginal generator is more expensive than in the \ssc case, as is the case for our case study. In other words, the minimization of total generation cost does not necessarily lead to the lowest average market prices. This is also seen for the month January for the 2100 MW case (green). However, in most cases and months the \mpa model does lead to lower average market prices compared to the \ssc model. 

As expected, the average market price decreases under increasing excess heat penetration. With a relatively high installed excess heat capacity of 2100 MW (green), the market price for the \mpa model is close to zero from April up to and including October. In the case of high installed capacity, the average marginal price difference between the two models is most pronounced. 


\section{Conclusions and Future Perspectives}\label{sec:conclusion}
We have investigated the consequences of integrating excess heat in district heating systems under two different scheduling and pricing paradigms: \ssc and \mpa. 
The \ssc is attractive due to its simplicity for both market operator and excess heat producers, but may lead to suboptimal scheduling. 
Our main conclusion is that at higher excess heat penetration, a simple price signal is no longer adequate, and more sophisticated pricing signals and/or other market setups may be needed.  
In our case study, we have shown that the disadvantages of using a price signal under high excess heat penetration include:
\begin{enumerate}
    \item \textbf{Expensive scheduling}: Excess heat is scheduled in hours where CHPs can produce relatively cheap heat instead of where CHPs produce more expensive heat. As a result, total CHP generation cost is higher under \ssc than the \mpa. 
    \item \textbf{Wasted excess heat}: Excess heat production is not matched to heat load, so that a greater amount of excess heat is  wasted.
    \item \textbf{High market prices}: Even though \mpa may lead to higher average market prices in some cases, market prices are most of the time higher in the \ssc model, especially under high excess heat penetration.
\end{enumerate}


\subsection{Discussion}
Under increasing penetration of excess heat, market prices decrease under both paradigms, and most drastically under \mpa. Market prices get close to zero during the summer time already for intermediate penetration of excess heat. This may affect the recovery of fixed cost for generators. For example, in \cite{Liu2019MarginalCost} it is shown that market revenues can be insufficient for investment cost recovery for most heat producers in a case study in the Netherlands. 
This problem has also been encountered in electricity systems with high shares of solar and wind energy \cite{Taylor2016PowerSystems}. For this case, it has been suggested that marginal-cost-based market clearing is not necessarily a proper solution for systems of generators with high fixed and low marginal costs, and a rethinking of power markets is needed \cite{Taylor2016PowerSystems}. 
This problem can be expected to arise in excess heat based heat systems too. 

We have concluded that more sophisticated pricing signals than the Stockholm price are needed in systems with high excess heat penetration. However, the Stockholm ambient-temperature dependent pricing signal has two attractive properties: transparency and interpretability. These properties should be considered when designing new methods for generating pricing signals. 

Finally, we note that cooling-based excess heat \prds provide most excess heat during the summer months, which is a mismatch with the load that is minimal in this period. This relation indicates that a seasonal storage may be a suitable supplement in systems dominated by cooling-based excess heat \prds.

\subsection{Recommendations for future work}
In this work, we have designed a model of cooling-based excess heat \prds with the aim was to mimic general dynamics of such \prds in a convex manner. The model has been formulated after discussion with experts in more detailed heat pump modeling. However, the model has not been verified using real data of excess heat \prds. This should be done in future work. 

Furthermore, the model could be extended and made more realistic in several ways. 
In our model, flexibility of heat producers was limited using an energy budget to be respected over every six hours. This is a stylized representation of excess heat flexibility. Future reformulations of the model could focus on improving the representation of this (limited) flexibility. 
\added{On the conventional generator side, the market considered here assumes that unit commitment constraints are included in the price-quantity bids. However, we have ignored unit commitment considerations in the construction of supply bids. Our work could be extended with block bids to represent bidding behavior of CHPs more realistically. }

\added{It could be argued that a scaling is needed to adapt the Stockholm heat price signal to the Copenhagen case. A sensitivity analysis for the value of scaling factor could show how this would affect our results. }

\added{We have addressed effects of excess heat integration on the day-ahead market, without considering potential sources of uncertainty. Future work could investigate how uncertainty of heat load, as well as uncertainty in excess heat production, could affect the scheduling and pricing of excess heat. 
We have also disregarded network considerations. Inclusion of such constraints could change our results quantitatively, as there would be a delayed arrival of CHP heat, while local excess heat would be delivered close to real time. In addition, heat transported from a distance would be accompanied by greater heat losses. Our work could be extended to include network constraints, for example using the linear formulation in \cite{Frolke2022Network}.}

Our analyses could also be extended by adding different types of market participants. For example, future work could investigate the effect of adding flexible loads to the system. It is furthermore likely that future district heating systems will include other excess heat \prds that are not cooling based, such as energy intensive industries. 




\section*{Acknowledgment}
This work is partly supported by EMB3Rs (EU H2020 grant no. 847121). 
The authors would like to thank Wiebke Meesenburg and Torben Schmidt Ommen from DTU Mechanical Engineering for discussion of convex heat pump modeling, and for the supply of temperature dependent COP profiles. We thank Tore Gad Kjeld of HOFOR for a constructive correspondence and kind provision of data, and Pierre Pinson for his inputs on an early version of this work. Finally, we thank two anonymous reviewers for their valuable feedback.

\appendix
We derive the price bids and feasible regions $\mathcal{F}$ for CHPs and excess heat producers in Appendices \ref{ap:CHPmodel} and \ref{ap:eh_model}, respectively.

\section{CHPs}\label{ap:CHPmodel}
Our model of CHP plants, including their bidding behavior, is identical to the sequential decoupled formulation in \cite{Mitridati2020Heat}. 
As it is the current practice in Copenhagen, we assume sequential heat and electricity markets, where the heat market is cleared first.

The fuel intake $F_{it}$ of CHP $i$ is equal to the fuel used for electricity generation $G\up{E}_{it}$ and heat generation $G\up{H}_{it}$. The fuel generation is upper bounded as
\begin{align}\label{eq:fuelbound}
    & F_{it} = \rho_i\up{E} G\up{E}_{it} + \rho_i\up{H}G\up{H}_{it} \leq \overline{F}_i  && \forall i,t  \,.
\end{align} 
Recall that $\rho\up{E}_i$ and $\rho_i\up{H}$ represent the fuel efficiency for electricity and heat, respectively. 
A minimum power-to-heat ratio $r_i$ relates heat and electricity production as
\begin{align}\label{eq:phratio}
    G\up{E}_{it} \geq r_{i} G\up{H}_{it}  && \forall i,t  \,.
\end{align}
Both heat and electricity generation must be non-negative, but due to \eqref{eq:phratio} limiting heat generation suffices. 
\begin{align}\label{eq:nonnegG}
    & 0 \leq G\up{H}_{it} \,.
\end{align}
If drawn in a diagram with $G\up{E}_{it}$ on the y-axis and $G\up{H}_{it}$ on the x-axis, constraints \eqref{eq:fuelbound}-\eqref{eq:nonnegG} form a triangle with the y-axis as a base. At the tip of the triangle, the amount of generated heat is 
\begin{align}
    G\up{H,*} = \frac{\overline{F}_i}{\rho\up{H}_i + r_i \, \rho\up{E}_i}  && \forall i \,.
\end{align}
Additionally, the heat generation of a CHP may be upper bounded as 
\begin{align}\label{eq:GHub}
    & G\up{H}_{it} \leq \overline{G}\up{H}_i && \forall i,t  \,.
\end{align}
As the heat market is cleared before the electricity market, constraints \eqref{eq:fuelbound}-\eqref{eq:GHub} can be replaced by the following bounds on the generated heat quantity:
\begin{align}
    0 \leq  G\up{H}_{it} \leq \min \left \{\overline{G}\up{H}_i, \frac{\overline{F}_i}{\rho\up{H}_i + r_i \, \rho\up{E}_i}  \right \}  && \forall i,t \,. 
\end{align}
Recall that we have already provided the feasible space for CHPS resulting from this constraint in Section \ref{sec:bidding} in Equation \eqref{eq:chp_feasible}.

The net heat production cost for a CHP is given by the difference of fuel cost and revenue from electricity sale:
\begin{align}
    C\up{H}_{it} = \alpha_i \, (\rho_i\up{E} G\up{E}_{it} + \rho_i\up{H} G\up{H}_{it} ) - \lambda_t\up{E} G_{it}\up{E} && \forall i,t \,,
\end{align}
where $\alpha$ is the fuel price. 
In \cite{Mitridati2020Heat}, the optimal heat bid $c\up{H}_{it}$ is derived for CHPs in a sequential heat and electricity market setting. The price bid depends on the (forecasted) electricity price as 
\begin{align}
    c\up{H}_{it} = 
    \begin{cases}
        \alpha_i \, (\rho\up{E}_i \,r_i + \rho\up{H}_i) - \lambda\up{E}_t \, r_i & \text{if } \lambda\up{E}_t \leq \alpha_i \rho_i\up{E} \\
        \lambda\up{E}_t \, \frac{\rho\up{H}_i}{\rho\up{E}_i} & \text{if } \lambda\up{E}_t > \alpha_i \rho\up{E}_i \,.
    \end{cases}
\end{align}



\section{\Wh \prds}\label{ap:eh_model}
We assume that all excess heat is produced as a by-product of cooling. In particular, we assume these agents cool their refrigeration cabinets using a local heat pump. 
The heat pump's heat output $G\up{H}_{et}$ relates to its electricity load $L\up{E}_{et}$ as 
\begin{align}\label{eq:ex1}
    G\up{H}_{et} = \cop_{et} L\up{E}_{et}   && \forall e,t \,,
\end{align}
where the coefficient of performance $\cop_{et}$ is a time-varying parameter in our model. This allows us to include its approximate ambient temperature dependence. The heat output of the heat pump is subject to upper and lower bounds:
\begin{align}
    0 \leq G\up{H}_{et} \leq \overline{G}_e\up{H} && \forall e,t \,.
\end{align}

The refrigeration cabinets are the heat source of the heat pump. This implies that the available heat depends on the temperature dynamics in these cabinets. We model the refrigerator temperature $T\up{F}_{et}$ using a linear difference equation as 
\begin{align}
T_{et+1}\up{F} - T_{et}\up{F} = A_e \, (T_{et}\up{I} -  T_{et}\up{F}) - B_e \, (G_{et}\up{H} - L_{et}\up{E})  && \forall e,t \,,
\end{align}
where $T\up{I}$ is the indoor temperature in the supermarket. The parameters $A_e$ and $B_e$ may differ per excess heat provider, depending on the physical characteristics of the refrigerators. 
The refrigerator temperature is subject to bounds:
\begin{align}
    \underline{T}_{e}\up{F} \leq T_{et}\up{F} \leq \overline{T}\up{F}  && \forall i,t \,.
\end{align}
The average temperature of the refrigerator over chosen time periods $\mathcal{P}$ must also stay within pre-set limits:
\begin{align}
    T\up{F-}_e \leq \frac{1}{|P|} \, \sum_{t \in {P}} T_{et}\up{F} \leq T\up{F+}_e && \forall e, P \in \mathcal{P} \,,
\end{align}
which ensures that the refrigerator temperature will not be on lower or upper bounds for longer periods of time. The periods must be defined such that $\bigcup_{P \in \mathcal{P}} P = \mathcal{T}$, so that all time steps are part of at least one period. 
Finally, the heat output from the heat pump is subject to ramping limits:
\begin{align}\label{eq:exend}
     \underline{R}_e \leq G\up{H}_{et+1} - G\up{H}_{et} \leq \overline{R}_e  && \forall e,t \,.
\end{align}
Recall that we already provided the feasible space for excess heat producers resulting from the previous constraints in Section \ref{sec:bidding} in Equation \eqref{eq:eh_feasible}.

\bibliographystyle{myIEEEtran}
\bibliography{setup/references_linde,setup/extra_references, setup/ieee_nourl}

\end{document}